\documentclass{ws-procs9x6}
\catcode`\@=11
\def\gsim{\ifmmode{\,\mathrel{\mathpalette\@versim>\,}}
    \else{$\,\mathrel{\mathpalette\@versim>}\,$}\fi}
\def\lsim{\ifmmode{\,\mathrel{\mathpalette\@versim<\,}}
    \else{$\,\mathrel{\mathpalette\@versim<}\,$}\fi}
\def\@versim#1#2{\lower 2.9truept \vbox{\baselineskip 0pt \lineskip
    0.5truept \ialign{$\m@th#1\hfil##\hfil$\crcr#2\crcr\sim\crcr}}}
\catcode`\@=12
\newcommand{\gv}{{\bf g}}

\newcommand{\xv}{{\bf x}}
\newcommand{\Sv}{{\bf S}}

\newcommand{\Mstar}{{M_*}}

\newcommand{\rstar}{r_*}

\newcommand{\az}{a_0}

\newcommand{\phiN}{\phi_{\rm N}}

\def\sglos{\sigma_{\rm los}}

\newcommand{\vstarn}{v_{\rm *n}}

\newcommand{\tdyn}{t_{\rm dyn}}
\newcommand{\tstard}{t_{\rm *d}}
\newcommand{\tstarn}{t_{\rm *n}}

\def\Re{R_{\rm e}}

\newcommand{\vstard}{v_{\rm *d}}
\newcommand{\Estard}{E_{\rm *d}}
\newcommand{\Estarn}{E_{\rm *n}}

\newcommand{\be}{\begin{equation}}
\newcommand{\ee}{\end{equation}}

\def\dxcube{d^3{\bf x}}

\def\Nr{N_{r}}

\def\Nth{N_{\vartheta}}
\def\Nph{N_{\varphi}}

\def\Tr{{\rm Tr}\,}

\def\vr{v_r}

\def\rc{r_{\rm c}}

\begin{document}

\title{PHASE MIXING IN MOND}

\author{L. CIOTTI$^*$ and C. NIPOTI}

\address{Dept. of Astronomy, University of Bologna,\\
via Ranzani 1, I-40127, Bologna, Italy\\
$^*$E-mail: luca.ciotti@unibo.it}

\author{P. LONDRILLO}

\address{INAF - Bologna Astronomical Observatory,\\
via Ranzani 1, I-40127, Bologna, Italy}

\begin{abstract}

Dissipationless collapses in Modified Newtonian Dynamics (MOND) have
been studied\cite{NLC07} by using our MOND particle-mesh N-body code,
finding that the projected density profiles of the final virialized
systems are well described by Sersic profiles with index $m \lsim 4$
(down to $m\sim 2$ for a deep-MOND collapse). The simulations provided
also strong evidence that phase mixing is much less effective in MOND
than in Newtonian gravity. Here we describe ``ad hoc" numerical
simulations with the force angular components frozen to zero,
thus producing radial collapses.  Our previous findings are confirmed,
indicating that possible differences in radial orbit instability under
Newtonian and MOND gravity are not relevant in the present context.

\end{abstract}


\bodymatter

\section{Introduction}

In the Lagrangian formulation of Milgrom's Modified Newtonian Dynamics
(MOND)\cite{BM84 ,M83} the Poisson equation $\nabla^2\phiN=4\pi G\rho$
for the Newtonian potential $\phiN$ is replaced by the field
equation for the MOND potential $\phi$
\begin{equation}
\nabla\cdot\left[\mu\left(\Vert\nabla\phi\Vert /\az\right)
                 \nabla\phi\right] = 4\pi G \rho,
\label{eqMOND}
\end{equation}
where $\rho$ is the density distribution, $\az\simeq 1.2 \times
10^{-10} {\rm m\, s^{-2}}$ is a characteristic acceleration, $\Vert
...\Vert$ is the standard Euclidean norm, and in finite mass systems
$\nabla\phi\to 0$ for $\Vert\xv\Vert\to\infty$.  The MOND
gravitational field experienced by a test particle is
$\gv=-\nabla\phi$, and $\mu(y)\sim y$ for $y\ll 1$ and $\sim 1$ for
$y\gg 1$ (typically $\mu =y/\sqrt{1+y^2}$).  In the `deep MOND regime'
describing low-acceleration systems ($\Vert\nabla\phi\Vert \ll\az$,
hereafter dMOND), $\mu(y)=y$ and so Eq.~(\ref{eqMOND}) simplifies to
$\nabla\cdot\left({\Vert\nabla\phi\Vert}\nabla\phi\right) = 4\pi G \az
\rho$. The source term in Eq.~(\ref{eqMOND}) can be eliminated by
using the Poisson equation, giving
\begin{equation}
\mu(\Vert\nabla\phi\Vert/\az)\nabla\phi=\nabla\phiN+\Sv,
\label{eqcurl}
\end{equation}
where $\Sv$ is a solenoidal field dependent on $\rho$ and in general
different from zero; when $\Sv=0$ Eq.~(\ref{eqcurl}) can be solved
explicitly in terms of $\nabla\phiN$. This reduction would be most
useful for numerical simulations. Unfortunately $\Sv =0$ only for very
special\cite{BM84, BM95} configurations.  In addition, though the
field $\Sv$ has been shown to be small\cite{BM95 ,CLN06} for some
configurations, neglecting it when simulating time-dependent dynamical
processes has dramatic effects such as non-conservation\cite{F84} of
the total linear momentum.

Several astronomical observational data appear consistent\cite{M02
,SMcG02} with the MOND hypothesis, and also a relativistic
version\cite{B04} of MOND is now available, making it an interesting
alternative to the cold dark matter paradigm.  However, dynamical
processes in MOND have been investigated very little\cite{BM99, BM00,
SK05, NP06, TC07} so far, mainly due to difficulties posed by the
non-linearity of Eq.~(\ref{eqMOND}). In a recent paper\cite{NLC07}
(hereafter NLC) we presented the results of N-body simulations of {\em
dissipationless collapse} in MOND obtained with our N-body code which
solves Eq.~(\ref{eqMOND}) exactly.  In particular, we obtained clear
indications that {\em phase mixing} is much less effective in MOND
than in Newtonian gravity. Here, after summarizing the main results of
NLC, and restricting for simplicity to the Newtonian and dMOND regimes
only, we address the problem of the importance of the force angular
components in the relaxation process, by running ``ad hoc" numerical
simulations with the force angular components frozen to zero.

\section{The N-body code and the numerical simulations}
\label{seccod}

Our MOND N-body code\cite{CLN06, NLC07} is based on a particle-mesh
scheme with quadratic spline interpolations.  The spherical grid on
which Eq.~(\ref{eqMOND}) is solved is made of
$\Nr\times\Nth\times\Nph$ points. We use leap-frog time integration,
where the adaptive time-step is the same for all particles.  All the
computations on the particles and the particle-mesh interpolations can
be split among different processors, while the iterative potential
solver computations are not performed in parallel: however, at each
time step we can use the potential previously determined as seed
solution.  We succesfully tested the code by running Newtonian
simulations (i.e., by solving Eq.~[\ref{eqMOND}] in the limit
$\mu=1$), and comparing the results with those of simulations
performed with our FVFPS\cite{LNC03, NLC03} treecode starting from the
same initial conditions.  We also verified that the code reproduces
the Newtonian and MOND conservation laws. In fact, $2K + W=0$ for
virialized systems in MOND and in Newtonian gravity, where $K$ is the
total kinetic energy and $W=\Tr W_{ij}$ is the trace of the potential
energy tensor\cite{BT87, C0}
\begin{equation}
\label{eqwij}
W_{ij}\equiv-\int\rho(\xv)x_i\frac{\partial\phi(\xv)}{\partial x_j}\dxcube .
\end{equation}
Note that in MOND $K+W$ is {\it not} the total energy, and is not
conserved. However, {\it $W$ is conserved in the limit of dMOND},
being $W(t)=-(2/3)\sqrt{G\az M_*^3}$ for {\it all} systems\cite{NLC07,
M84, GS92, M94} of finite total mass $\Mstar$.

The choice of appropriate scaling physical units is an important
aspect of MOND N-body simulations.  A full discussion of this point
can be found in NLC; here we just recall that, while in Newtonian
simulations the natural scaling units are $\tstarn=\rstar^{3/2} (G
\Mstar)^{-1/2}$, $\vstarn=(G \Mstar)^{1/2}\rstar^{-1/2}$, and
$\Estarn=G \Mstar^2\rstar^{-1}$, in the dMOND case one has
$\tstard=\rstar (G \Mstar\az)^{-1/4}$, $\vstard=(G \Mstar \az)^{1/4}$,
and $\Estard=(G\az)^{1/2}\Mstar^{3/2}$, where $\rstar$ and $\Mstar$
are the length and mass units in which the initial conditions
are expressed.
\begin{figure}
\psfig{file=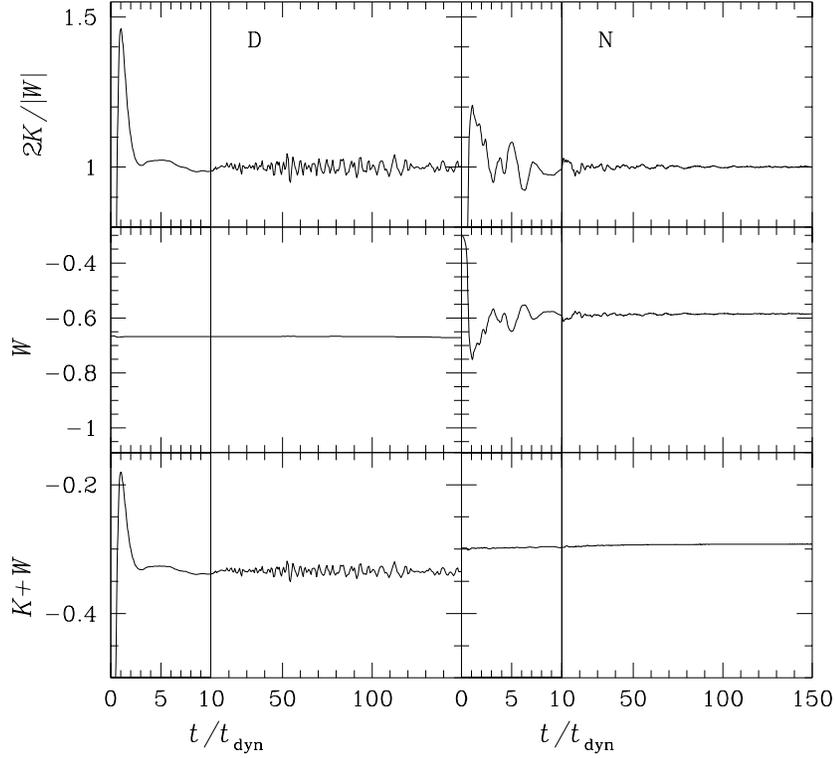,width=4.3in}
\caption{Time evolution of $2 K/|W|$, $W$, and $K+W$ for
         a dMOND (D) and a Newtonian (N) simulation. $K$, $W$, and $K+W$ 
         are in units of $\Estard$ (D), and $\Estarn$ (N).
         Note the long-lasting oscillations of the virial ratio in D.}
\label{figtime}
\end{figure}
In NLC we performed a set of N-body simulations of dissipationless
collapses, starting from the same phase-space configuration, i.e.,
from a cold ($2K/|W|=0$) Plummer\cite{P11} sphere of total mass
$\Mstar$ and ``core" radius $\rstar$.  The gravitational potential is
Newtonian in simulation N and dMOND in simulation D; the full MOND
simulations are not described here (see NLC).  All the simulations
($N=10^6$ particles, $\Nr=64$, $\Nth=16$ and $\Nph=32$) are evolved up
to $t=150\tdyn$, where $\tdyn$ is defined as the time at which
$2K/|W|$ reaches its maximum value ($\tdyn\sim 2\tstard$ in simulation
D, and $\sim 2\tstarn$ in N).  The center of mass, and the
modulus of the total angular momentum (in units of
$\rstar\Mstar\vstarn$ for model N and of $\rstar\Mstar\vstard$ for
model D) oscillates around zero with r.m.s $\lsim 0.1\rstar$ and
$\lsim 0.02$, respectively.  The quantities $K+W$ in simulation N, and
$W$ in simulation D are conserved to within $2\%$ and $0.6\%$,
respectively (see Fig.1).  The final angle-averaged density profiles
are fitted with a $\gamma$-model\cite{D93, T94}
\begin{equation}
\label{eqgamma}
\rho (r)=\frac{(3-\gamma)\Mstar \rc}{4\pi r^{\gamma} (\rc +r)^{4-\gamma}};
\end{equation}
for each end-product we also measure the axis ratios $c/a$ and $b/a$
of the inertia ellipsoid\cite{NLC02, NLC06} ($a\geq b\geq c$), the
ellipticity $\epsilon$ of the projections along the principal axis,
and the corresponding circularized effective radius $\Re$.  The resulting 
circularized projected density profiles are fitted
with the Sersic\cite{S68, C91} law
\begin{equation}
\label{eqser}
I(R)=I(\Re)\,
{\rm e}^{-b(m)\left[(R/\Re)^{1/m} -1 \right]},
\end{equation}
where\cite{CB99} $b(m)\simeq 2m-1/3+4/405m$.  Note that $m$ is the
only free parameter, because $\Re$ and $I(\Re)$ are fixed by particle
count.

\section{Results}
\label{secres}

\subsection{Structure and kinematics of the collapse end-products}

All the final virialized systems in NLC depart significantly from
spherical symmetry. In particular, the D end-product is triaxial
($c/a\sim0.2$, $b/a\sim0.4$; $0.5\lsim\epsilon\lsim 0.8$), while model
N is oblate ($c/a \sim c/b \sim 0.5$; $0\lsim\epsilon\lsim 0.5$).
These values are consistent with those observed in real ellipticals,
with the exception of model D, which would correspond to an E8 galaxy.
Thus, MOND gravity could be able to produce some system that would be
unstable in Newtonian gravity.  The final angle-averaged density of
model N is well described by Eq.~(\ref{eqgamma}) with $\gamma\sim
1.7$, while $\gamma\sim 0$ in model D.  Equation~(\ref{eqser}) fits
remarkably well the final surface density profiles (best-fit index
$m\sim 4$ in model N, and $m\sim2$ in D, see Table 2 in NLC), with
average residuals $0.05 \lsim\langle\Delta{SB}\rangle \lsim 0.2$,
where $SB\equiv-2.5 \log [I(R)/I(\Re)]$. Note that the fitting radial
range $0.1\,\lsim\, R/\Re\,\lsim\, 10$ is comparable with or larger
than the typical ranges\cite{BCD02} spanned by observations.  The
internal kinematics of the end-products is quantified by the
angle-averaged radial and tangential components of their
velocity-dispersion tensor ($\sigma_r$ and $\sigma_{\rm t}$), and by
the anisotropy parameter $\beta(r) \equiv 1 -0.5\sigma^2_{\rm
t}/\sigma^2_r$.  All systems are strongly radially anisotropic outside
the half-mass radius.  The $\sigma_r$ profile decreases steeply in the
final state of model N, while it presents a hole in the inner regions
of model D. In addition, model D is radially anisotropic
($\beta\sim0.4$) even in the central regions, where model N is
approximately isotropic ($\beta \sim 0.1$).  The line-of-sight
velocity dispersion $\sglos$ declines steeply within $\Re$ in model N,
while the D profile is significantly flatter.

\subsection{Phase-space properties}
\label{secphsp}

In Newtonian gravity, collisionless systems virialize through violent
relaxation in few dynamical times, as predicted by the
theory\cite{LB67} and confirmed by numerical\cite{vA82, NLC06}
simulations. Due to the non linearity of the theory, the details of
relaxation processes and virialization in MOND are much less known.
In Fig.~\ref{figtime} we show the time evolution of $2 K/|W|$, $W$,
and $K+W$ for simulations D and N of NLC.  In simulation N, $2 K/|W|$
has a peak, then oscillates, and eventually converges to the
equilibrium value 1; the total energy $K+W$ is nicely conserved.  The
time evolution is significantly different in simulation D, where $W$
is constant as expected, but $2 K/|W|$ {\it still oscillates at very
late times} because of the oscillations of $K$.
\begin{figure}
\psfig{file=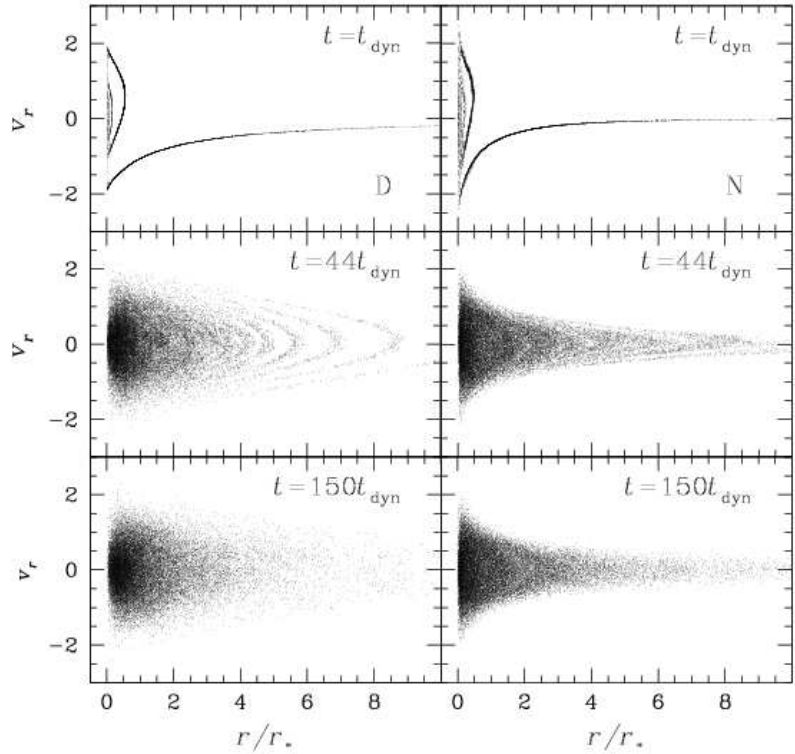,width=4.3in} 
\caption{Radial-velocity vs. radius of 32000 particles 
         randomly extracted from simulations D and N.
         $\vr$ is in units of $\vstard$ (D) or $\vstarn$ (N).}
\label{figphsp}
\end{figure}
A different view of phase space is given in Fig.~\ref{figphsp}, where
we plot time snapshots of the particles radial velocity vs. radius for
simulations D and N. At $t=\tdyn$ (time of the peak of $2 K/|W|$),
sharp shells in phase space are present, indicating that particles are
moving in and out collectively and phase mixing has not taken place
yet.  At significantly late times ($t=44\tdyn$), when the systems are
almost virialized ($2 K/|W|\sim 1$), phase mixing is complete in
simulation N, but phase space shells still survive in model D.
Finally, the bottom panels show the $(r,\vr)$ plane at equilibrium
($t=150\tdyn$), when phase mixing is completed also in model D: note
how the populated regions are significantly different for the two
models. Thus, NLC results indicate that phase mixing in more effective
in Newtonian gravity than in MOND. {\it Here we address the issue of
the importance of the force angular components during the collapse. In
fact, one could speculate that a different behavior of radial orbit
instability in MOND and in Newtonian gravity could be at the origin of
the different time scale of phase mixing}. As shown in Fig.3, 
also the new simulations confirm that phase mixing is less effective
in MOND than in Newtonian gravity (even though virialization times are
now longer than in NLC models, due to the reduced number of active
degrees of freedom).

In NLC we obtained additional information on the relaxation process
from the differential energy distribution\cite{BT87} $N(E)$ (i.e. the
number of particles with energy per unit mass between $E$ and $E+dE$).
In Newtonian gravity $\phi$ is usually set to zero at infinity for
finite-mass systems, so $E=v^2/2+\phi(\xv) <0$ for bound particles; in
MOND all particles are bound independently of their velocity, because
$\phi$ is confining, and all energies are admissible (see Fig.~5 of
NLC).  Given that the particles are at rest at $t=0$, the initial
$N(E)$ depends only $\phi(\xv)$ at $t=0$, and it is significantly
different in the Newtonian and MOND cases.  In accordance with
previous studies, we found that in the Newtonian case the final $N(E)$
is well represented by an exponential function\cite{B82, vA82, C91,
LMS91, TBVA05, NLC06} over most of the populated energy range.  In
contrast, in model D the final $N(E)$ decreases for increasing energy,
qualitatively preserving its initial shape.  {\it We interpret this
result as another manifestation of a less effective phase space
reorganization in MOND than in Newtonian collapses}.
\begin{figure}
\psfig{file=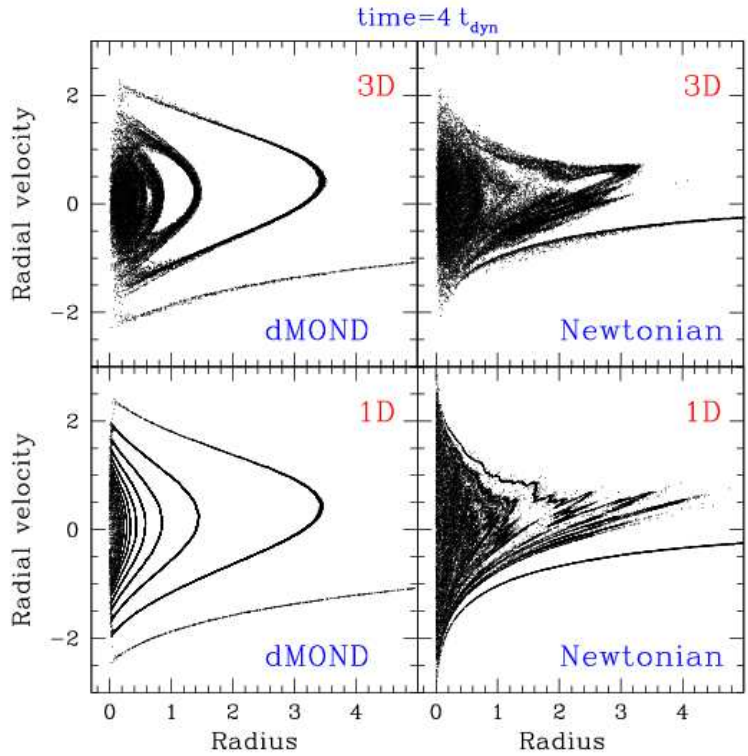,width=4.3in} 
\caption{Phase space sections for simulations with 
         frozen angular force components (bottom panels), and with the 
         three force components active (top panels).}
\label{figrad}
\end{figure}

\section{Conclusions}
\label{secdis}

We presented results of dissipationless collapses in MOND, focusing on the
relaxation process. The main results can be summarized as follows:

1) Newtonian collapses produced cuspier density profiles than MOND
simulations ($\gamma\sim 1.7$ and $m\sim 4$ vs. $\gamma\sim 0$ and
$m\sim 2$). In both cases the Sersic fits are remarkably good over a
large radial interval. In addition, Newtonian models are isotropic in
the central regions, while dMOND models are radially anisotropic down
to the center.

2) In NLC we found that final states of full MOND models, if
interpreted in the context of Newtonian gravity, are characterized by
a {\it dividing radius} of the order of $\Re$, separating a
baryon-dominated inner region from a dark-matter dominated outer
region, in accordance\cite{B94, C06} with observations of elliptical
galaxies. However, we were not able to reproduce the observed scaling
laws of elliptical galaxies {\it under the assumption of a
luminosity-independent stellar mass-to-light ratio}.

3) Phase mixing is less effective (and stellar systems take longer to
relax) in MOND than in Newtonian gravity. This behavior is confirmed
by numerical simulations in which the angular force components are
frozen to zero, so possible differences in radial orbit instability
between Newtonian and MOND gravity are not relevant in the present
context.  Our results on mixing suggest that merging could take longer
in MOND than in Newtonian gravity; on the other hand, analytical
estimates of the two-body relaxation time seem to indicate\cite{CB04}
the opposite, predicting shorter dynamical friction time-scales in
MOND than in Newtonian gravity: the next application of our code will
be the study of galaxy merging in MOND.

\section*{Acknowledgments}

We are grateful to Giuseppe Bertin, James Binney, and Scott Tremaine
for helpful discussions, and to Italian MIUR for the grant CoFin2004.

\end{document}